# Reconfigurable Acoustic Metalens with Tailored Structural Equilibria

Dinh Hai Le[1,2,*], Felix Kronowetter[2], Yan Kei Chiang[1], Marcus Maeder[2], Steffen Marburg[2], David A. Powell[1,†]

[1] School of Engineering and Technology, University of New South Wales, Northcott Drive, Canberra, ACT 2600, Australia.

[2] Chair of Vibroacoustics of Vehicles and Machines, Department of Engineering Physics and Computation, TUM School of Engineering and Design, Technical University of Munich, Garching b., 85748 München, Germany.

Email: [*]hai.le@unsw.edu.au and [†]david.powell@unsw.edu.au

## ABSTRACT

The ability to concentrate sound energy with a tunable focal point is essential for a wide range of acoustic applications, offering precise control over the location and intensity of sound pressure maxima. However, conventional acoustic metalenses are typically passive, with fixed focal positions, limiting their versatility. A significant obstacle in achieving tunable sound wave focusing lies in the complexity of precise and programmable adjustments, which often require intricate mechanical or electronic systems. In this study, we present a theoretical and experimental investigation of a reconfigurable acoustic metalens based on a bistable origami design. The metalens comprises eight flexible origami units, each capable of switching between two stable equilibrium states, enabling local modulation of sound waves through two distinct reflection phases. The metalens can be locked into specific symmetric or asymmetric configurations by manually tailoring the origami units to settle either of the two states. Each configuration generates a unique phase profile, focusing sound energy at a specific point. This concept allows the focal spot to be dynamically reconfigured both on and off-axis. Furthermore, the approach introduces a simple yet effective mechanism for tuning sound energy concentration, offering a solution for flexible acoustic manipulation.

# INTRODUCTION

In recent decades, metasurfaces—engineered subwavelength structures—have undergone rapid and transformative development, impacting many fields, including elastics[1–4], seismics[5,6], electromagnetics[7–10], and acoustics[11,12]. One particularly significant application of metasurfaces is their use for focusing wave energy, commonly known as metalenses. These metalenses have proven essential for enhancing various technologies, such as imaging[13–15], communication systems[16], energy harvesting[17–19], and biomedical devices[20,21].

In the acoustic domain, metalenses leverage local or non-local local phase modulation[22–24] to focus and manipulate sound waves. Acoustic metalenses have emerged as effective tools for addressing several limitations inherent in traditional acoustic lenses, such as narrow frequency ranges[25], fixed focal points[26], sensitivity to the angle of incidence[23], low energy concentration efficiency[22], and limited reconfigurability[27]. However, a major challenge with most existing acoustic metalenses is their rigidity, which restricts their application in dynamic environments where adjustable focal points or varying focal lengths are required. This limitation usually stems from fabricating these metalenses from a single material with fixed physical properties. One common solution to overcome this challenge involves using actuator-driven structures. However, these systems are often complex, expensive, and require high power consumption[26–29]. Another approach involves manually arranging and rearranging unit cells to achieve specific configurations for different focusing behaviors, which, although practical, is time-consuming[25,30,31]. As a result, single-material acoustic metalenses may struggle to achieve advanced functionalities such as multi-focalization, phase-gradient control, or adaptive focusing, which are essential for more versatile and dynamic applications.

Building on the Generalized Snell's Law (GSL) design approach, this study addresses the limitations of fixed focalization and poor reconfigurability in single-material acoustic metalens by introducing a novel self-locking deformable metalens capable of switchable focal spot location. This concept builds upon our recent work on multi-material 3D-printed metasurfaces composed of identical bistable origami units used for 1-bit coded acoustic manipulation, enabling reconfigurable beam focusing and splitting[32]. In this study, we extend that approach to develop a reconfigurable acoustic metalens, offering enhanced precision and control over the focusing behavior. The metalens consists of eight multi-material 3D-printed origami units, which are identical in form but differ in size. Each origami unit possesses two equilibrium states, enabling modulation of two distinct reflection phase shifts for normally incident impinging sound at 2000 Hz. As a result, the combined configuration of eight origami units facilitates 16 different phase responses, ensuring full $2\pi$ phase coverage. By reconfiguring the metalens into eight distinct configurations, each characterized by a unique combination of first and second equilibrium states, eight focal points can be adjusted along both on-axis and off-axis locations. We validate the switchable focusing capabilities of the metalens through numerical simulations, theoretical analysis and experimental demonstrations. The reconfigurable behavior results from the inherent flexibility of the origami units, enabled by advanced multi-material 3D-printing techniques. This approach represents a significant advancement in developing highly reconfigurable and multifunctional metasurfaces, offering straightforward tuning and a self-locking mechanism.

# RESULTS

## Origami metalens approach

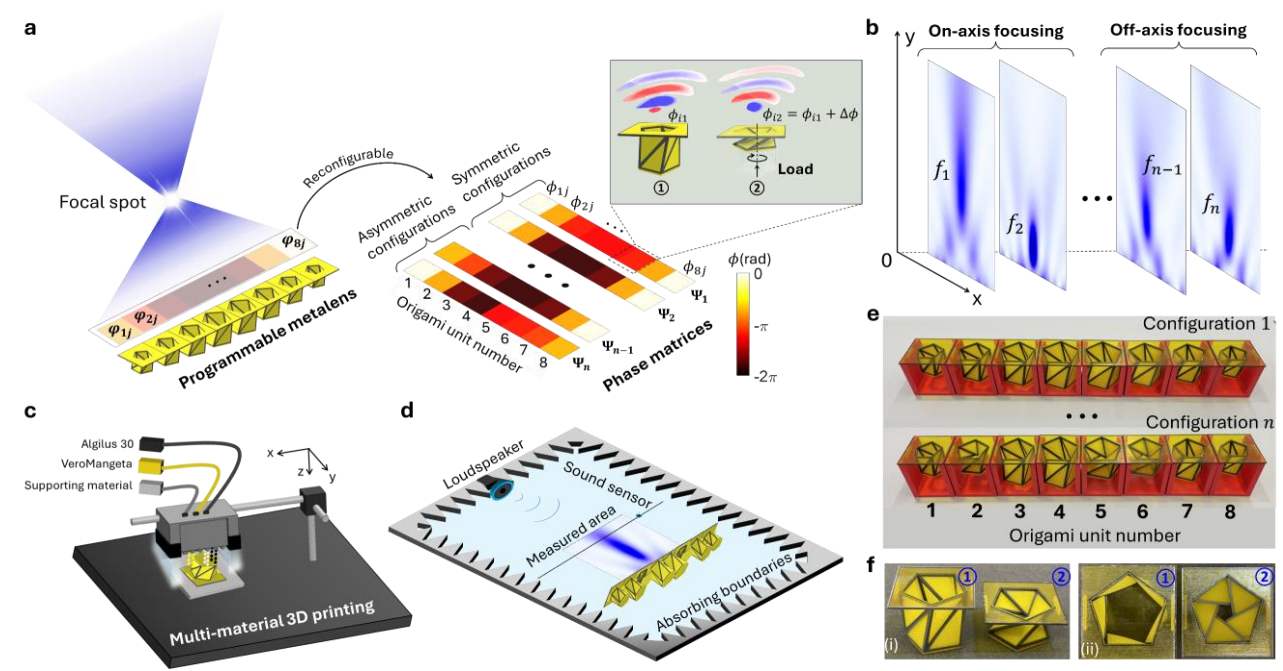

**Figure 1. Illustration of reconfigurable origami metalens concept**. **a**) Programmable acoustic metalens with adjustable focal spots. The metalens can be switched into multiple distinct configurations, each characterized by a unique reflection phase matrix $\Psi_n$ (where $n = 1, 2, \ldots, 256$ denotes the configuration number). The inset depicts the concept of *i*-th bistable origami unit (where $i = 1, 2, \ldots, 8$), the first ($j = 1$) and second ($j = 2$) equilibrium states, which modulate different reflection phase shifts, $\phi_{i1}$ and $\phi_{i2} = \phi_{i1} + \Delta\phi_i$. **b**) Controllable focussing, where each focal spot corresponds to a specific phase profile, enabling programmable focalization of acoustic energy. **c**) Schematic illustration of the multi-material printing technology. **d**) Measurement setup for the realization of metalens. **e**) Fabricated one-dimensional metalens in two different arbitrary configurations, with red frames supporting the origami units during measurement. **f**) Fabricated origami unit in its first (left) and second (right) equilibrium states, shown in (i) perspective and (ii) top views. The yellow color represents rigid, while the dark grey indicates flexible materials.

In the following, we provide the concept of the proposed acoustic metalens (Figure 1a), in which we developed a one-dimensional (1D) metalens to focus acoustic energy at 2000 Hz. The metalens consists of eight bistable deformable unit cells inspired by a non-rigid Kresling origami pattern[33], derived from four distinct structures that are identical in form but differ in size. Specifically, we use identical pairs for units 1 and 8, 2 and 7, 3 and 6, and 4 and 5. Units 1 and 8 are the smallest, while units 4 and 5 are the largest. The origami unit settles in its first or second equilibrium state without external interaction. If all the origami units settle in the same equilibrium state, the metalens exhibits perfect symmetry. This symmetry can be disrupted by selectively tailoring the deformation state of individual unit cells. The switching behavior in each origami unit results from applying a left- or right-handed twist of the torque, $\boldsymbol{T}(\gamma)$ with an angle $\gamma$, or an axial Force, $\boldsymbol{F}(h)$. It results in a height change, $\Delta h$, of the origami unit. Under normal incident sound, the $i$-th origami unit (where $i = 1, 2, ..., 8$) modulates the reflection phase value of $\phi_{ij}$. Here, $j = 1$ represents the phase at the first equilibrium, while $j = 2$ denotes the phase at the second equilibrium. This design allows each origami unit to function as a switchable phase modulator with a phase shift of $\phi_{i1}$ and $\phi_{i2} = \phi_{i1} + \Delta\phi_i$ (Figure 1b). It demonstrates that the 8-unit metalens will modulate 16 different phase profiles. Given that each of the eight origami units can switch between two equilibrium states, the metalens theoretically reproduces 256 unique configurations, each producing a different 1D reflection phase profile $\boldsymbol{\Psi}_n = [\phi_{1j}\ \phi_{2j}\ ...\ \phi_{8j}]$ (where $n = 1, 2, ..., 256$ and $j = 1, 2$). We investigate the focusing behavior of the metalens across four symmetric and four asymmetric configurations, resulting in eight distinct on- and off-axis focal points. To experimentally validate the focusing function of the metalens, we perform measurements in a parallel plate waveguide, as illustrated in Figure 1d. The scattered sound field is extracted from the total acoustic pressure field using a $k$-space filtering technique (see the Methods Section for more details). Figure 1e exhibits the fabricated 1D metalens. Each origami unit originates from a multi-material 3D printing with Polyjet technology, which allows for the simultaneous deposition of two materials with different mechanical properties (Figure 1c). Specifically, we chose Agilus30, a rubber-like material, for the soft hinges, while the rigid compound VeroMagenta builds the rigid walls. The details of the fabrication process follow in the Methods Section.

## Design principle

The key to focusing sound waves lies in constructive interference at a specific point in space. This concept requires that the wavefront reflected from the metalens interferes constructively at a focal point by satisfying the focusing condition of the GSL[34]:

$$\phi_i = 2\pi n - k(r_i - f_i), \tag{1}$$

where $\phi_i$ denotes the reflection phase of the $i$-th origami unit, $k = 2\pi/\lambda$ is the wave number with $\lambda$ being the wavelength of the sound wave. Furthermore, $r_i$ is the $i$-th unit from the origin, $f_i$ represents the corresponding distance to the focal point, and $n$ is an integer ensuring constructive interference occurs. Each unit settles into one of two stable states, corresponding to a different reflection phase. The reflection phase for these two equilibria must satisfy[23]:

$$\phi_1(x_i) = -k \cdot \frac{{x_i}^2}{2f_1}; \quad \phi_2(x_i) = -k \cdot \frac{{x_i}^2}{2f_2}. \qquad (2)$$

Here, $x_i \in [-3.5P, 3.5P]$ is the position of the $i$-th origami unit, $P$ is the dimension of the origami units. In addition, $f_1$ and $f_2$ are the focal lengths corresponding to the two equilibrium states. The phase shifts across the metalens must follow a parabolic distribution to ensure constructive interference and to focus on the desired focal point. For two adjacent units $i$ and $i+1$, the phase differences in the first and second equilibria are:

$$\Delta\phi_{i1} = \phi_{(i+1)1} - \phi_{i1} = k \cdot \frac{P(2x_i + P)}{2f_1}, \qquad (3)$$

$$\Delta\phi_{i2} = \phi_{(i+1)2} - \phi_{i2} = k \cdot \frac{P(2x_i + P)}{2f_2}. \qquad (4)$$

The critical factor that allows for switching between focal points is the phase difference between the two stable states, $\Delta\phi(x_i)$, given by:

$$\Delta\phi(x_i) = \phi_1(x_i) - \phi_2(x_i) = \frac{k{x_i}^2}{2}\left(\frac{1}{f_2} - \frac{1}{f_1}\right). \qquad (5)$$

Using this framework, we designed the origami units of the metalens to exhibit a gradient phase shift in both equilibrium states. Furthermore, we tailor the reflection phase profile for each unit to follow a parabolic distribution, ensuring constructive interference and focusing on the designed focal points.

## Origami units and mechanical properties

The origami unit design follows the Kresling origami pattern, derived from a flat sheet configuration as depicted in Figure S1a, Supplementary Note 1. It comprises ten identical triangular panels ABC connected by valley and mountain creases. In the folded state, the $N$-sided origami exhibits one or two equilibrium states depending on its structural parameters, including the number of sides $N$, side length $a$, and the angles $\alpha$ and $\beta$. For this study, we chose $N = 5$ (pentagon), $\alpha = 50°$, and $\beta = 32°$ to ensure bistable characteristics, while the side length $a$ is varied to optimize the desired designs. Furthermore, three parameters characterize the folded configuration, i.e., the radius $r$, the height $h$, and the twist angle $\gamma$. The origami unit is a bi-material structure consisting of a rigid wall, shown in yellow, and flexible hinges, depicted in dark grey (Figure S1b, Supplementary Note 1). The configuration consists of three layers, including a fixed top squared surface with thickness $t_2$, the origami layer, with wall thickness $t_1$, and the bottom rotatable surface, also with thickness $t_2$, which is attached to the lower end of the origami layer. Detailed geometric parameters of an origami unit provides Table S1, Supplementary Note 1.

The proposed metalens comprises eight origami units optimized to fit the measurement setup. The symmetric arrangement of the metalens includes: Units 1 and 8 with $a = 24$ mm; Units 2 and 7 with $a = 30$ mm; Units 3 and 6 with $a = 34$ mm; and Units 4 and 5, positioned at the center of the metasurface, with $a = 38$ mm. Figure 2a and Table S2, Supplementary Note 1, present the geometric parameters of the eight origami units.

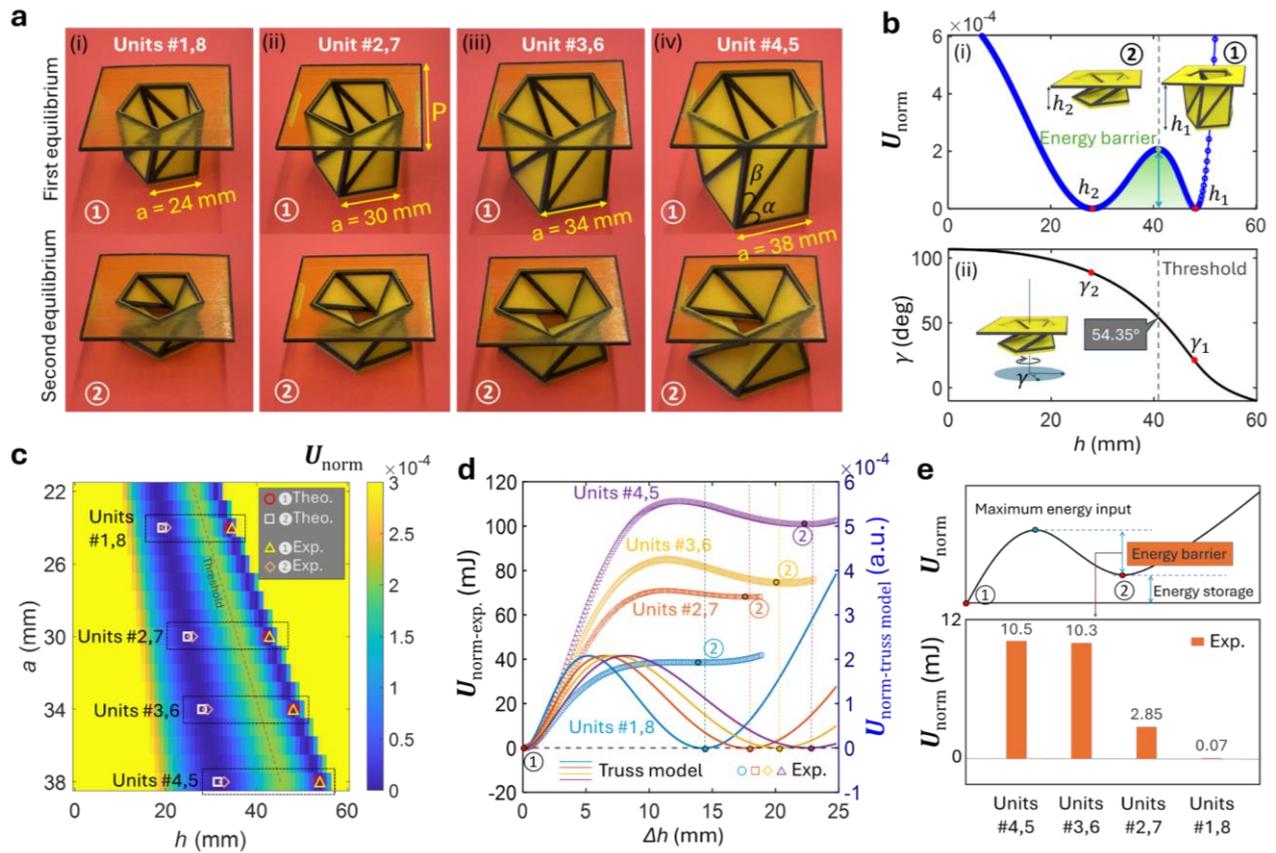

**Figure 2. Design of the bi-material origami unit and mechanical properties**: **a**) Prototypes of the origami units: (i) Units 1 and 8; (ii) units 2 and 7; (iii) units 3 and 6, and (iv) units 4 and 5, with each unit capable of remaining in either first (denoted as ①) or second (denoted as ②) equilibrium. **b**) Theoretical analysis of (i) normalized strain energy and (ii) twist angle of the origami unit as functions of height $h$ for side length $a = 32$ mm. **c**) Normalized strain energy and the corresponding positions of the two equilibrium states as functions of $a$, with the scatter points representing the positions of the two equilibria obtained from both theory and experiment. **d**) Comparison of normalized strain energy between the truss model and experimental results during displacement, $\Delta h$. **e**) Comparison of measured energy barriers among origami units.

The deformation of the origami unit involves coupled vertical translation and twisting, characterized by height $h$ and twist angle $\gamma$, respectively. This deformation occurs within the origami layer. By applying an axial force $\boldsymbol{F}(h)$ or exerting a torque $\boldsymbol{T}(\gamma)$ on the bottom layer, the origami unit will undergo collapse or deploy transitions. The positions of the stable states result from applying the principle of minimum strain energy. A visualization of the normalized strain energy stored within the origami provides Figure 2b(i), calculated from the following equation based on an equivalent truss model[35]:

$$\boldsymbol{U}_{\text{norm}}(h,r,\gamma) = \frac{N}{2(L_{AB} + L_{BC} + L_{AC})} (L_{AB} \cdot \varepsilon_{AB}^2 + L_{BC} \cdot \varepsilon_{BC}^2 + L_{AC} \cdot \varepsilon_{AC}^2), \tag{6}$$

where $L_{AB}, L_{BC}, L_{AC}$ are the length of line $AB, BC, AC$, and $\varepsilon_{AB}, \varepsilon_{BC}, \varepsilon_{AC}$ are the corresponding strains[32,35] along the lateral direction of each truss. In this analysis, the height, $h$, is considered a fixed parameter, while the radius $r$, and twist angle $\gamma$ are dependent variables. To determine the equilibrium states and establish the relationships among $r, h, \gamma$, and normalized strain energy $\boldsymbol{U}_{\text{norm}}$, different loading conditions are considered, such as axial deformation with free rotational movement and torsional loading with unconstrained lateral displacement. These conditions lead to solving the following nonlinear equations:

$$\left.\frac{\partial \boldsymbol{U}_{\text{norm}}(h,r,\gamma)}{\partial \gamma}\right|_h = \left.\frac{\partial \boldsymbol{U}_{\text{norm}}(h,r,\gamma)}{\partial r}\right|_h = 0,$$
$$\left.\frac{\partial^2 \boldsymbol{U}_{\text{norm}}(h,r,\gamma)}{\partial \gamma^2}\right|_h > 0, \quad \left.\frac{\partial^2 \boldsymbol{U}_{\text{norm}}(h,r,\gamma)}{\partial r^2}\right|_h > 0, \tag{7}$$

where $\partial \boldsymbol{U}_{\text{norm}}(h,r,\gamma)/\partial r|h$ and $\partial \boldsymbol{U}_{\text{norm}}(h,r,\gamma)/\partial \gamma|h$ denote the restoring force $F$ and torque $T$, respectively, during the deformation process. Figure 2b(ii) presents the relation between the twist angle and height of one origami unit. The results indicate that the first equilibrium state appears at a height of 44 mm with a twist angle of 19.5°, and the second equilibrium state is obtained at 21 mm when the twist angle is 88.5°. The threshold twist angle defines the transition point between two equilibria, determined to be 54.35°. The origami unit will be either collapsed or deployed to transform back and forth between two equilibria by left- or right-handed twisting the bottom layer. Figure 2c illustrates the dependence of the two equilibrium positions on the side length $a$ with the scatter points representing the positions of the two equilibria obtained from both theory and experiment. The results indicate that, as $a$ increases from 22 mm to 38 mm, the first and second equilibrium points rise almost linearly from 34 mm to 53.8 mm and from 19.2 mm to 31.4 mm, respectively. The measured heights at the first and second equilibrium points agree well with the truss-model analysis.

To quantify the origami units' energy barrier and deformation time, we assessed their mechanical behavior under axial compression using the Instron Tensile Tester 5942, as shown in Figure S4a, Supplementary Note 4. See the Methods Section for further details of the experimental setup. Figure S4b, Supplementary Note 4 shows the force measurements over displacement, $F(\Delta h)$, for four origami units. Units with higher values of parameter $a$ generated stronger maximum restoring forces. Specifically, units 1 through 4 reached maximum restoring forces of $F(\Delta h)_1 = 7.5$ N, $F(\Delta h)_2 = 12.3$ N, $F(\Delta h)_3 = 13.9$ N, and $F(\Delta h)_4 = 16.4$ N, respectively. Figure 2d compares the normalized strain energy between the theoretical truss model and experimental results as a displacement function, $\Delta h$. The results of normalized strain energy under the applied compression result from integrating the force over the displacement as $U_{\text{norm}}(\Delta h) = \int F(\Delta h) d(\Delta h)$. It is worth noting that the difference in the unit of $U_{\text{norm}}$ between the truss model and the measurements is due to the factor, $EA$, the axial rigidity of the truss elements. This discrepancy arises from the complex stretching and folding stiffness of the actual structures, which the truss model does not account for, as it omits structural thickness and material properties[36]. The second equilibrium positions in the truss model show slight deviations compared to the experimental measurements. Furthermore, while the theoretical energy at the second equilibrium state drops to zero, the measured energy remains above zero. This discrepancy arises from the elastic deformation of the flexible hinges, which undergoes both bending and stretching. Notably, origami units with larger values of $a$ exhibit higher strain energy at the second equilibrium state due to their greater elastic volume. The difference between the maximum energy input and the strain energy at the second equilibrium is the energy barrier (Figures 2a and 2c), indicating the energy required to transition the structure between two equilibrium states. If the energy barrier exceeds zero, the origami unit remains bistable; if it falls below zero, it becomes monostable. We compared the measured energy barriers of the origami units shown in Figure 2e, which demonstrate a rapid decrease in energy barrier as the size of the origami units decreases. Specifically, units 4 and 5 have an energy barrier of 10.5 mJ, while units 1 and 8 exhibit a significantly lower energy barrier of 0.07 mJ in the experiment. The range of $a$ values, between 24 mm and 38 mm, is chosen to comply with the maximum allowable height of 68 mm in the measurement setup while ensuring that the strain energy barrier remains greater than zero.

## Acoustic phase profiles for reconfigurable configurations

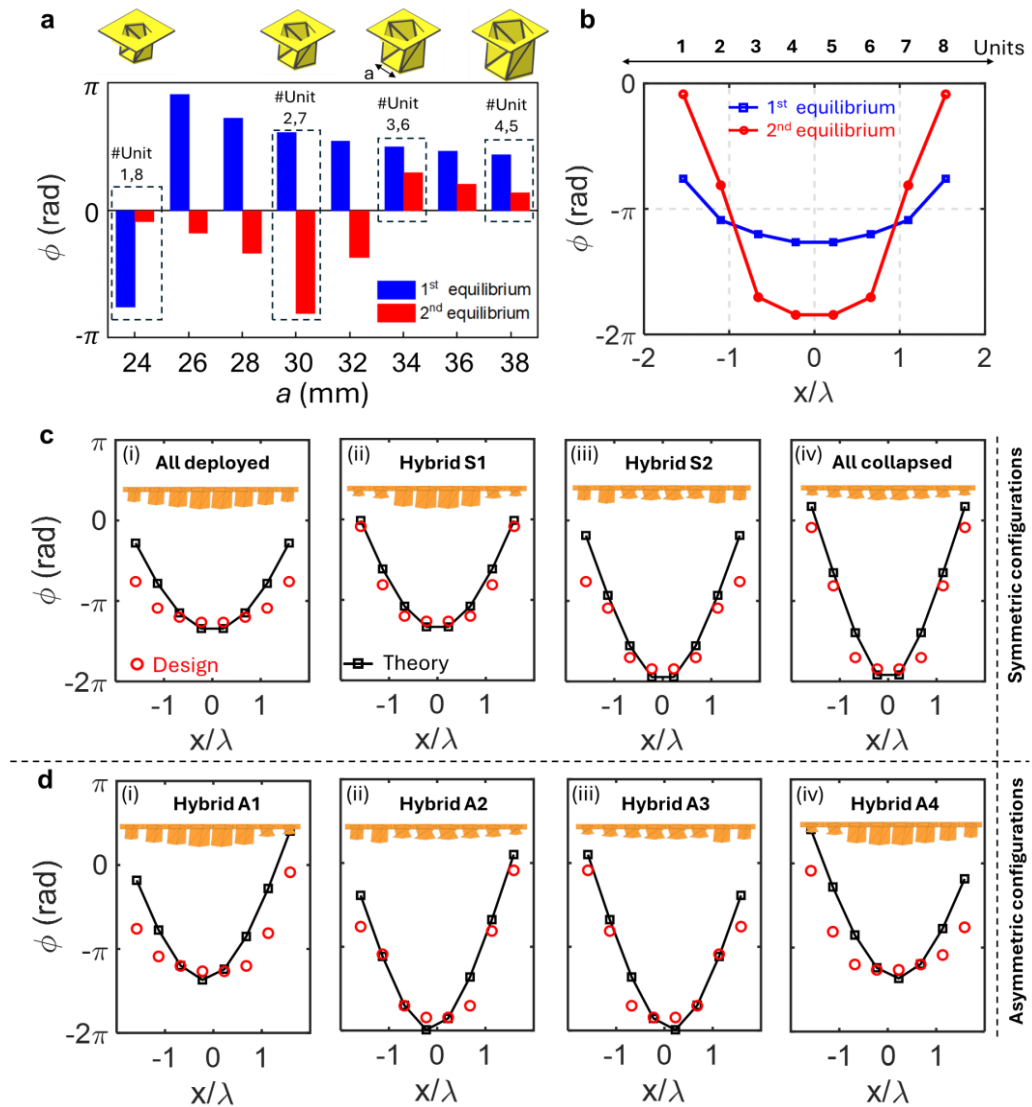

**Figure 3. Reflection phases of the origami units and various configurations of the metalens**. **a**) Reflection phase as a function of $a$ for the two equilibria. Specific values of $a = 24, 30, 34,$ and $38$ mm are chosen for units 1, 8, units 2, 7, units 3, 6, and units 4, 5, respectively. **b**) The reflection phase distribution of the metalens at the first and second equilibrium points. **c**) The tailored phase distribution of symmetric configurations of the metalens: (i) all origami units remain in the first stable state ("All Deployed"); (ii) Units 1, 2, 7, and 8 stay at the first equilibrium, while units 3, 4, 5, and 6 transition to the second equilibrium ("Hybrid S1"); (iii) Reverse of configuration Hybrid 1 ("Hybrid S2"); and (iv) all origami units transition to the second stable state ("All collapsed"). **d**) The reproduced phase distribution of asymmetric configurations of the metalens: (i) Units 1–6 remain in the first equilibrium, while units 7 and 8 transition to the second equilibrium ("Hybrid A1"); (ii) Units 1 and 2 stay at the first equilibrium, while units 3–8 transition to the second equilibrium ("Hybrid A2"); (iii) Reverse of configuration Hybrid A2 ("Hybrid A3"); and (iv) Reverse of configuration Hybrid A1 ("Hybrid A4"). Red circles denote the simulated reflection phase shift of the designed metalens, and the black squares are the calculated phase shifts according to the GSL theory. Insets depict the configurations of metalens corresponding to each case.

The results of the simulated reflection phase of the origami unit as a function of the parameter $a$ provides Figure 3a. Since the reflection phase depends on the origami height[32], which corresponds to the propagation path length of the acoustic wave, an increase in origami height leads to a shift in the reflection phase at both equilibria. The results demonstrate that the reflection phase can be modulated over the full $2\pi$ range. As the maximum value of $a$ is 38 mm, we selected the origami with $a = 38$ mm as units 4 and 5, which have the reflection phases at first and second equilibria, are $\phi_1(x_4) = \phi_1(x_5) = -3.9$ rad and $\phi_2(x_4) = \phi_2(x_5) = -5.9$ rad. For the remaining units, the reflection phases must satisfy the phase differences outlined in Equations (3-5), with compensatory phase shifts added to the reflection phases in both equilibria. The compensatory phase shifts at first, $\Delta\phi_{\text{comp1}}$, and at the second equilibrium, $\Delta\phi_{\text{comp2}}$, ensure that the overall phase profile maintains the required parabolic shape for focusing:

$$\Delta\phi_{\text{comp1}} = \phi_1(x_4) - \frac{k}{2f_1}x_4^2,$$

$$\Delta\phi_{\text{comp2}} = \phi_2(x_4) - \frac{k}{2f_2}x_4^2. \tag{8}$$

The phase for all remaining units $i = 1, 2, \ldots, 8$ is calculated using the general parabolic profile for each stable state as:

$$\phi_1(x_{i \neq 4,5}) = k \cdot \frac{x_i^2}{2f_1} + \Delta\phi_{\text{comp1}},$$

$$\phi_2(x_{i \neq 4,5}) = k \cdot \frac{x_i^2}{2f_2} + \Delta\phi_{\text{comp2}}. \tag{9}$$

The predicted phase profile of the metalens before compensation (as described by Equation (2)) and after compensation (see Equation (9)) highlights Figure S2, Supplementary Note 2. Based on this analysis, the design parameters for the origami units are chosen from Figure 3a as follows: $a = 24$ mm for units 1 and 8; $a = 30$ mm for units 2 and 7; and $a = 34$ mm for units 3 and 6. Figure 3b illustrates the relative phase modulations of the 8-unit metalens in the first (blue line) and second equilibrium states (red line). In the first equilibrium state, the reflection phase of the metalens ranges from $-2.2$ rad to $-3.9$ rad, while in the second equilibrium state, it spans from $-0.3$ rad to $-5.9$ rad. This configuration ensures that the metalens maintains the desired switchable focusing behavior, enabling constructive interference at two distinct focal points depending on the stable state of the unit cells.

Interestingly, the 16 discrete phase shifts can be systematically reproduced into four symmetric and four asymmetric 8-unit phase profiles, resulting in eight distinct configurations of the proposed metalens. In the symmetric configurations shown in Figure 3c, Pure P1 (all deployed) involves all origami units in the first stable state. In Hybrid S1, units 1, 2, 7, and 8 remain in the first equilibrium, while units 3, 4, 5, and 6 transition to the second. Hybrid S2 is the reverse of Hybrid S1, with units 3, 4, 5, and 6 in the first equilibrium and units 1, 2, 7, and 8 in the second. In Pure P2 (all collapsed), all units transition to the second stable state. For the asymmetric configurations illustrated in Figure 3d, Hybrid A1 has units 1–6 in the first equilibrium and units 7 and 8 in the second. In Hybrid A2, units 1 and 2 remain in the first equilibrium while units 3–8 transition to the second. Hybrid A3 is the reverse of Hybrid A2, with units 3–8 in the first equilibrium and units 1 and 2 in the second. Hybrid A4 reverses Hybrid A1, with units 7 and 8 in the first equilibrium and units 1–6 in the second. These eight configurations generate eight distinct focal spots. Additional configurations of the metalens are discussed in Supplementary Note 7.

## Switchable metalens

The focusing performances of the proposed metalens in its eight different configurations are numerically and experimentally demonstrated at 2000 Hz, as illustrated in Figure 4. Appropriate simulations realized the adjustable focusing in these configurations, with excellent agreement confirmed in an experimental measurement. We designed the simulation setup to demonstrate the focusing behavior of the metalens to replicate the dimensions of the parallel plate waveguide used in the measurement across all $x$, $y$, and $z$ axes (see details in Methods Section). The results validate our concept of multi-focusing with deep focus using a programmable origami metalens. In the symmetric configurations, the metalens enables adjustable on-axis focusing along the vertical axis (Figure 4a), ranging from $f_1 = 360$ mm in the All-Deployed mode to $f_2 = 100$ mm in the all-collapsed mode. The Hybrid S1 and S2 configurations focus sound energy at positions of $f_3 = 250$ mm and $f_4 = 160$ mm, respectively. The corresponding focal spots observed in the experiment are measured at 380 mm, 260 mm, 165 mm, and 105 mm, as shown in Figure 4b. On the other hand, with the asymmetric configuration exhibited in Figure 4c, the metalens facilitates off-axis focusing with focal spots varying in both x and y directions. Specifically, the Hybrid A1 and A2 configurations focus acoustic energy at the focal spot of $f_5(-34 \text{ mm}, 240 \text{ mm})$ and $f_6(-28 \text{ mm}, 120 \text{ mm})$, respectively, whereas the Hybrid A3 and A4 configurations focus at opposite positions along the x-axis at $f_7(34 \text{ mm}, 240 \text{ mm})$ and $f_8(28 \text{ mm}, 120 \text{ mm})$. The corresponding focal spots observed in the experiment are at $(-34 \text{ mm}, 240 \text{ mm})$, $(-34 \text{ mm}, 240 \text{ mm})$, $(-34 \text{ mm}, 240 \text{ mm})$, and $(-34 \text{ mm}, 240 \text{ mm})$, as shown in Figure 4d. To demonstrate the strong agreement between simulation and measurement of the focusing behavior, Figures 4e and 4f compare normalized acoustic intensities: on-axis focusing (solid line) and off-axis focusing (scatter plot). For on-axis focusing, intensities are extracted along a vertical cutline at the $x = 0$ axis, while for off-axis focusing, intensities are derived from horizontal cutlines at each focal plane. These results experimentally verify acoustic focusing on and off-axis tunability, confirming the theoretical predictions.

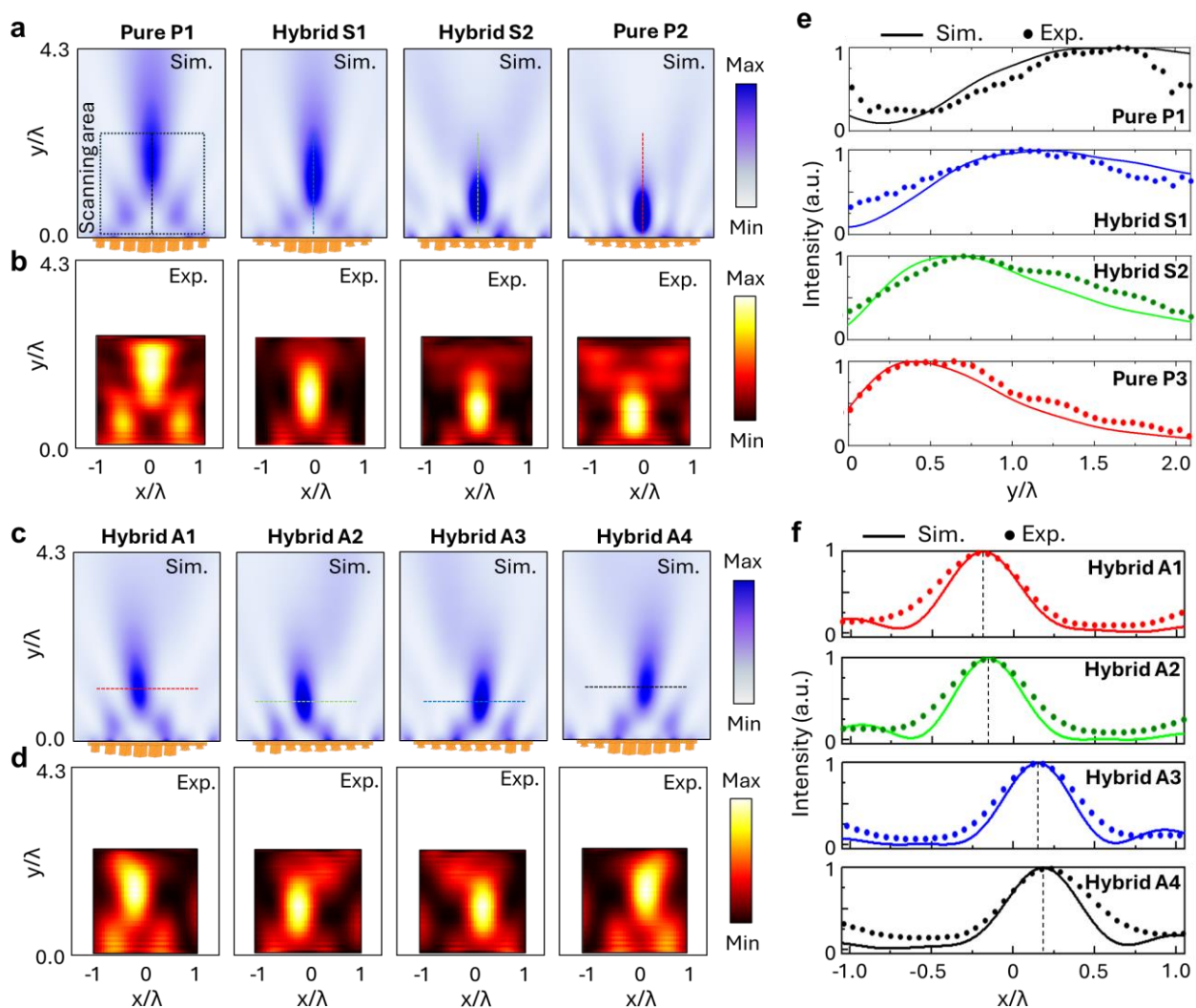

**Figure 4. Numerical and experimental verification of the metalens with reconfigurable focusing.** **a**) Simulated and **b**) measured acoustic intensity distributions of the metalens for on-axis focalizations. **c**) Simulated and **d**) measured acoustic intensity distribution of the metalens for off-axis focalizations. Normalized acoustic intensities are compared between simulation and measurement, with e) depicting the on-axis focus (solid line) and f) illustrating the off-axis focus (scatter plot).

## CONCLUSIONS

In conclusion, we have introduced an approach for programmable focalization of sound energy using an acoustic metalens, leveraging multi-material 3D-printing technology. The metalens comprises eight flexible origami units, each featuring a straightforward design with two equilibria. These origami units, fabricated via multi-material 3D printing, can lock into their first or second equilibrium states, thereby modulating the sound field through two distinct reflection phase profiles. We engineer the reflection phase shift of each origami unit to align with the Generalized Snell's Law for continuous phase delay. The 8-unit metalens, capable of generating 16 distinct phase profiles, can thus be reproduced into eight self-locking configurations, focalizing acoustic energy into eight on- and off-axis focal spots. Furthermore, we demonstrate the focusing performance of the metalens numerically and experimentally, revealing that the acoustic energy convergence can be adjusted on- and off-axis within an area of $0.2\lambda \times 2\lambda$ at an operating frequency of 2000 Hz, depending on whether the metalens is programmed in symmetric or asymmetric configurations.

Our straightforward design allows the development of multifunctional and highly reconfigurable devices with precise manipulation of acoustic wavefronts without relying on complicated control systems. This innovative concept can be extended to the electromagnetic domain, allowing for the creation of tunable, switchable, and programmable electromagnetic metasurfaces for diverse applications, for example, frequency selective surfaces, radar cross-section reduction, and polarization converters. As outlined in recent studies, some specific applications require active control of the origami units tuned using actuators[37], magnetic fields[38], heat[39], or electrothermal[40].

# METHODS.

## Numerical simulation

We conduct 3D finite element method-based simulations using the acoustic module of COMSOL Multiphysics 6.2. The background pressure field is a normal incident plane wave in the air medium with a mass density and a sound speed of $1.22 \text{ kg} \cdot \text{m}^{-3}$, and $343 \text{ m} \cdot \text{s}^{-1}$, respectively. The origami unit has a non-thickness design with sound-hard boundary conditions. Furthermore, perfectly matched layers (PML) prevent undesired reflections at the top boundary for simulations of a single origami unit and at both the top and side boundaries for metalens characterization. In addition, we design the simulation setup to align the dimensions of the parallel plate waveguide in the experimental measurements.

## Fabrication

The origami unit is multi-material 3D printed using Polyjet (material jetting) technology in the printer Stratasys J826 (Stratasys Ltd.), which features a resolution of 14 μm layer height and 40 μm in the $x$- and $z$-axes. A detailed schematic of the fabrication process of an origami unit provides Figure S3, Supplementary Note 3. The bimaterial used consists of Agilus, a rubber-like material with a Shore A hardness of 30 for the flexible hinges, and VeroMagenta V, with a Shore D hardness of 85, for the rigid walls. In addition, we employ a gel-like material (SUP706B) during the printing process and subsequently remove it using a Waterjet cleaning system. Finally, we let the structures air-dry at room temperature. The 3D support frames are fabricated from polylactic acid (PLA) using Fused Deposition Modeling (FDM) technology with a Prusa 3.9 3D printer.

## Compression measurement

We design an experiment to decouple both linear translation and rotational motion to validate the compression behavior of four individual origami units. The experimental setup utilizes the Instron Tensile Tester 5942 (Figure S4, Supplementary Note 4). Each origami unit is anchored at its base, while the top connects to a custom-designed, 3D-printed free-rotation joint. This joint is engineered to convert axial displacement into rotational motion and secures the origami unit using side tapes. During testing, we measure the restoring force with a force sensor featuring a maximum capacity of 20 N, an accuracy of $\pm 0.5\%$, and high resolution for detecting minute force variations. The compression rate is set to $0.01 \text{ mm} \cdot \text{s}^{-1}$. We manually adjust the force sensor for each test to the initial position corresponding to the first equilibrium point.

## Acoustic measurement

We measure the metalens in a parallel plate waveguide (Figure S5, Supplementary Note 5). Absorbing materials surrounding the waveguide, with a height of 68 mm, minimize reflections. Furthermore, a loudspeaker positioned 80 cm from the metalens sample generates a normal incident sound field. In addition, a microphone mounted on a movable belt, with point-by-point measurements taken at 1 cm intervals over an area of $36 \times 38 \text{ cm}^2$ ($2.1\lambda \times 2.2\lambda$) records the complex values of the total acoustic pressure. The measured total acoustic pressure, $P_{tot}(x,y)$, is a superposition of incident, $P_{inc}(x,y)$, and scattered fields, $P_{scat}(x,y)$, expressed as:

$$P_{tot}(x,y) = P_{inc}(x,y) + P_{scat}(x,y). \tag{10}$$

We employ the 2D Fast Fourier Transform to the total field to convert the measured data from the spatial domain to the wavenumber domain to extract the scattered pressure field from the total acoustic pressure as:

$$P(k_x, k_y) = \mathcal{F}\{P_{tot}(x,y)\} = \int_{-\infty}^{\infty} \int_{-\infty}^{\infty} P_{tot}(x,y) e^{-\mathrm{i}(k_x x + k_y y)} \mathrm{d}x \mathrm{d}y, \tag{11}$$

where i denotes the imaginary unit and $k_x$ and $k_y$ represent the wavenumbers in the corresponding spatial direction, respectively. The Fourier-transformed field $P(k_x, k_y)$ contains both the incident and scattered components. Here, we introduce three components of the $k$-space filtering function, $H_1(n), H_2(n)$ and, $H_3(n)$, to construct the complete filter for isolating the incident field, expressed as:

$$H_1(n) = \frac{1}{2}\left(1 + \frac{2}{\pi}\tan^{-1}(4n)\right), \tag{12a}$$

$$H_2(n) = -\frac{1}{2}\frac{2}{\pi}\tan^{-1}[4(n-n_{-1})] - \frac{1}{2}, \tag{12b}$$

and

$$H_3(n) = -\frac{1}{2}\frac{2}{\pi}\tan^{-1}[4(n-n_0)] + \frac{1}{2}, \tag{12c}$$

where $n$ represents discrete points, $n_{-1}$, $n_0$ correspondingly represent lower and upper cutoff indexes in $k_y$-space. The complete filtering function $H(k_y)$, is then given by combining these components:

$$H(k_y) = H_1(k_y) + H_2(k_y) + H_3(k_y). \tag{13}$$

Details of this filtering function provides Supplementary Note 6. By applying $H(k_y)$ as a filter in the k-space, we can isolate the incident component from the total field:

$$P_{ky,inc}(x, k_y) = H(k_y) \cdot P(k_x, k_y). \tag{14}$$

Subtracting the filtered incident component from the total field in $k_y$-space isolates the scattered component:

$$P_{scat}(k_x, k_y) = P(k_x, k_y) - P_{ky,inc}(x, k_y). \tag{15}$$

Furthermore, we obtain the scattered spatial domain field by using the inverse Fourier transform of the separated frequency-domain fields as:

$$P_{scat}(x, y) = \mathcal{F}^{-1}\{P_{scat}(k_x, k_y)\}. \tag{16}$$

This three-component filtering function enables a refined separation of the incident and scattered fields in the spatial domain, enhancing the accuracy of the scattered pressure field extraction.

# ACKNOWLEDGEMENTS

**Funding:** This research is supported by Australian Research Council Discovery Project DP200101708, and Universities Australia/DAAD joint research co-operation Scheme Project 57446203.

**Author contributions:** D. H. L: conceptualization, methodology, software, validation, investigation, data curation, writing (original draft), visualization. F. K: software, investigation, writing (review and editing). M. M: investigation, supervision, writing (review and editing). Y. K. C: methodology, validation, investigation, supervision, writing (review and editing). S. M: resources, writing (review and editing), supervision. D. A. P: conceptualization, investigation, resources, writing (review and editing), supervision, project administration, funding acquisition.

**Competing Interests:** The authors declare no competing financial interests.

**Code availability:** The code used for the analyses will be made available upon reasonable request to the corresponding author.

**Data and materials availability:** All data needed to evaluate the conclusions in the paper are present in the paper and/or the Supplementary Materials.

Supplementary Information for

# Reconfigurable Acoustic Metalens with Tailored Structural Equilibria

**Dinh Hai Le[1,2,*], Felix Kronowetter[2], Yan Kei Chiang[1], Marcus Maeder[2], Steffen Marburg[2], David A. Powell[1,†]**

[1] School of Engineering and Technology, University of New South Wales, Northcott Drive, Canberra, ACT 2600, Australia.

[2] Chair of Vibroacoustics of Vehicles and Machines, Department of Engineering Physics and Computation, TUM School of Engineering and Design, Technical University of Munich, Garching b., 85748 München, Germany.

Email: [*]hai.le@unsw.edu.au and [†]david.powell@unsw.edu.au

**This file includes:**

Figure S1-S5

Table S1-S2

Supplementary Note 1. Geometrical parameters of origami units

Supplementary Note 2. Predicted phase profile for switchable focusing function

Supplementary Note 3. Fabrication process of origami unit

Supplementary Note 4. Experimental verification for axial load test.

Supplementary Note 5. Measurement setup for metalens characterization

Supplementary Note 6. K-space filleting function

Supplementary Note 7. Acoustic behaviors of additional configurations

## Supplementary Note 1: Geometrical parameters of origami units

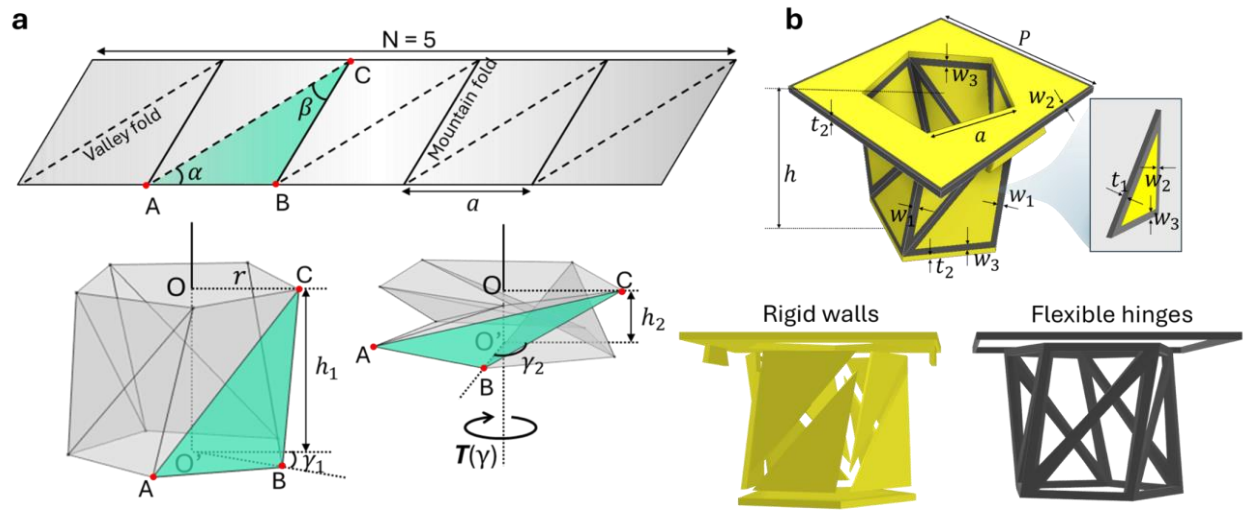

**Figure S1. Kresling origami and designed bi-material origami unit**: **a**) The Kresling origami pattern in its flat sheet and folded forms. **b**) Perspective view of the designed origami unit at its first equilibrium, illustrating the structural parameters and its composition of rigid walls and flexible hinges.

**Table S1**. Designed parameters of an origami unit cell at first equilibrium state.

| Parameters | $P$ (mm) | $a$ (mm) | $t_1$ (mm) | $t_2$ (mm) | $w_1$ (mm) | $w_2$ (mm) | $w_3$ (mm) | $h_1$ (mm) | $\alpha$ (deg) | $\beta$ (deg) |
|---|---|---|---|---|---|---|---|---|---|---|
| Values | 66 | 32 | 1.2 | 2.0 | 1 | 1.5 | 2.0 | 42.2 | 50 | 32 |

**Table S2. Geometric parameter of eight origami units.**

| Unit no. | $a$ (mm) | $r$ (mm) | $h_1$ (mm) | $h_2$ (mm) | $\gamma_1$ (deg) | $\gamma_2$ (deg) |
|---|---|---|---|---|---|---|
| #1, 8 | 24 | 20.4 | 34 | 19.2 | 19.5 | 88.5 |
| #2, 7 | 30 | 25.52 | 42.5 | 24.8 | 19.6 | 88.6 |
| #3, 6 | 34 | 28.92 | 48.2 | 28 | 19.6 | 88.6 |
| #4, 5 | 38 | 32.32 | 53.8 | 31.4 | 19.6 | 88.5 |

**Supplementary Note 2: Predicted phase profile for switchable focusing function**

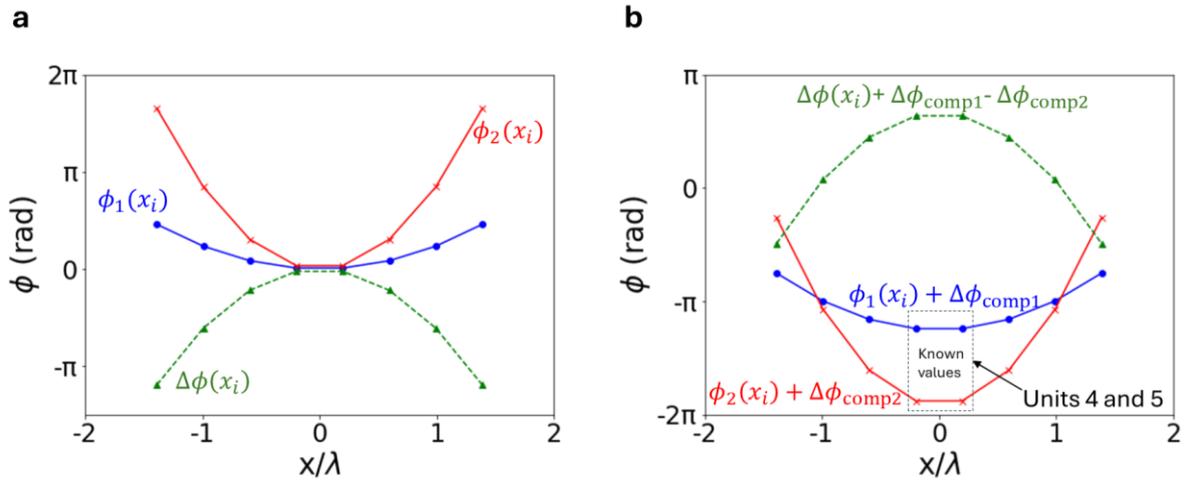

**Figure S2. Predicted phase profile for switchable focusing function**. a) Corresponds to Equations (2) and (5), and b) corresponds to Equation (9). The blue and red lines represent the phase profiles of the metalens when all units are in the first and second equilibrium states, respectively. The green line shows the phase difference for each unit.

**Supplementary Note 3: Fabrication process of origami unit**

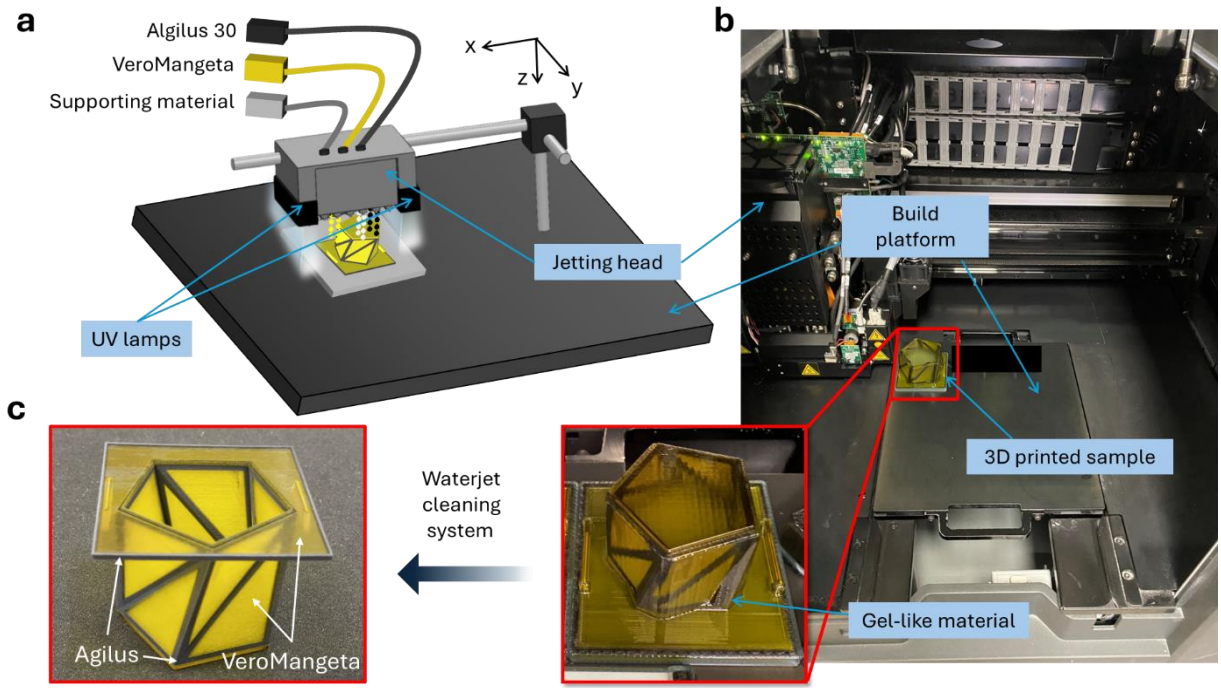

**Figure S3**. **Fabrication process of origami unit using multi-material 3D printing technology**. a) Schematic illustration of the multi-material printing technology. b) Photograph showing the internal components of the Stratasys J826 printer. c) Image of the fabricated origami unit.

**Supplementary Note 4. Experimental verification for axial load test.**

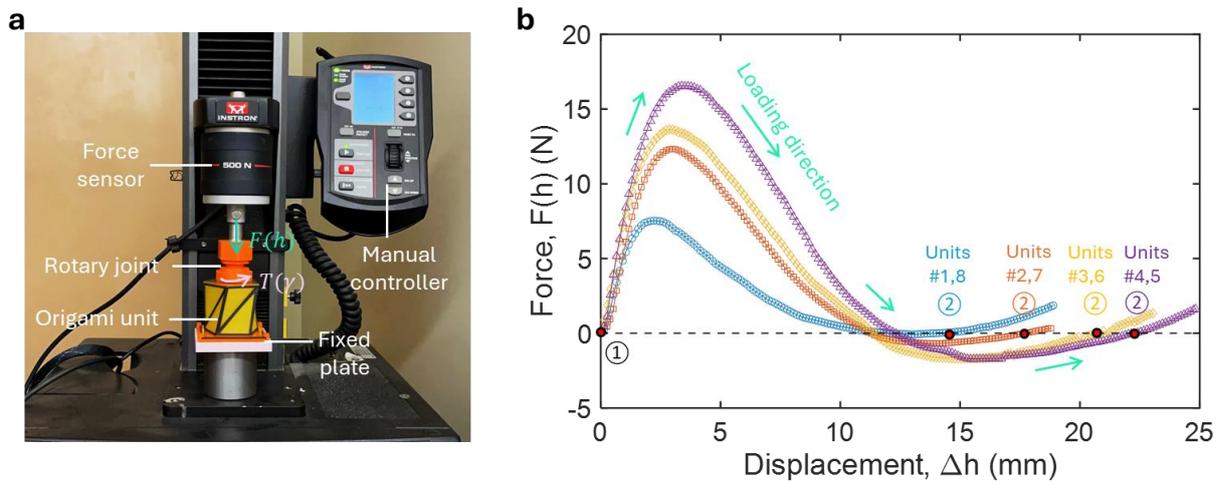

**Figure S4. Experimental verification for axial load test.** a) Experimental setup and b) measured results of axial loading versus displacement.

**Supplementary Note 5. Measurement setup for metalens characterization**

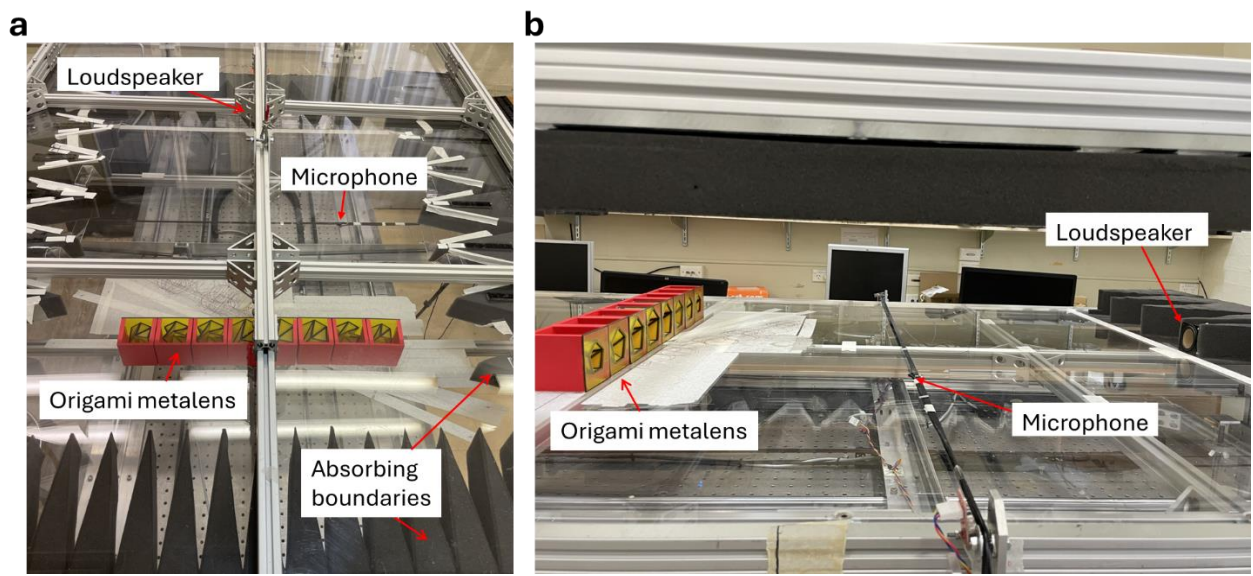

**Figure S5. Measurement setup for metalens characterization.** a) Top and b) side views.

**Supplementary Note 6. K-space filleting function**

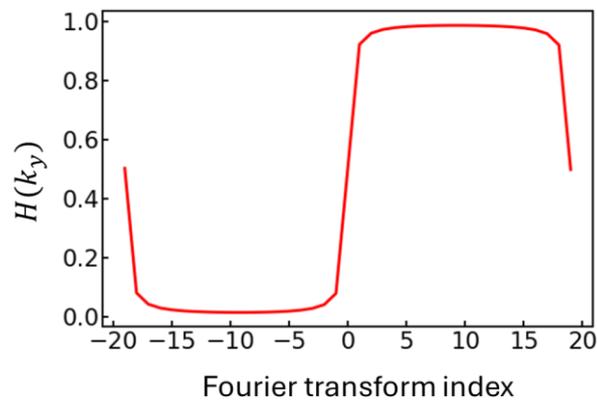

**Figure S6. K-space filleting function.** K-space filtering function $H(k_y)$, designed to isolate the incident and scattered field components in the total measured field. This filtering function, as illustrated in the figure above, serves as a band-pass filter in the spatial frequency domain along the $k_y$-axis. The construction of $H(k_y)$ is carefully designed with three components $H_1(k_y)$, $H_2(k_y)$, and $H_3(k_y)$ that together create a filter with a step-like response in $k_y$-space. The step-like behavior of the filter allows specific $k_y$-values associated with the incident field to pass while attenuating those associated with the scattered field. The smooth transitions at the edges, created by combining multiple arctan-based components, help reduce ringing artifacts in the spatial domain.

**Supplementary Note 7. Acoustic behaviors of additional configurations**

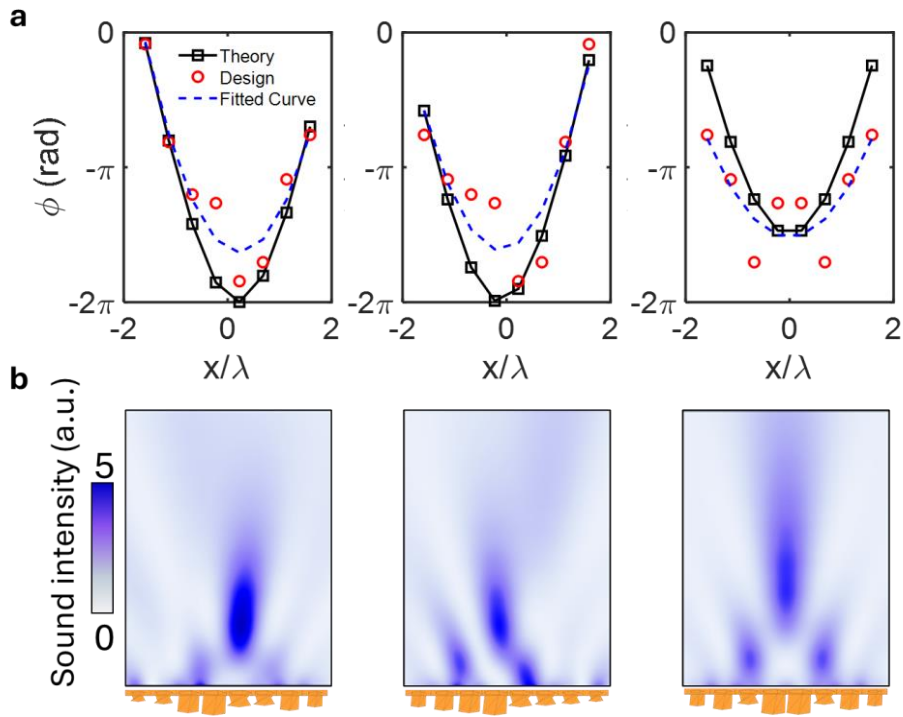

**Figure S7**. **Focusing behaviors of additional configurations**. (Top) The design phase profiles of three additional configurations and (Bottom) the corresponding sound intensity maps.

The proposed metalens offers a total of 256 possible configurations; however, not all of them exhibit focusing behavior. To achieve focus, the reflection phases of the origami units must satisfy equation (2). In some configurations, although the actual phase matrices do not perfectly align with the parabolic shape, their fitting curves generally follow a parabolic trend. We numerically demonstrate the focusing behavior of several additional configurations, as illustrated in Figure S6. The results show that the sound intensity of the metalens significantly decreases in configurations where the phase shifts deviate further from the theoretical phases required for focusing.